\documentclass[conference]{IEEEtran}
\usepackage{hyperref}
\usepackage{graphicx}
\usepackage{url}
\let\labelindent\relax
\usepackage{enumitem}
\usepackage{subcaption}
\usepackage{wrapfig}

\title{Using machine learning for medium frequency derivative portfolio trading}

\author{\IEEEauthorblockN{Abhijit Sharang}
\IEEEauthorblockA{Department of Computer Science\\
Stanford University\\
Email: abhisg@stanford.edu}
\and
\IEEEauthorblockN{Chetan Rao}
\IEEEauthorblockA{Stanford University, Netflix Inc.\\
Email: chetanr@stanford.edu}
}

\begin{document}

\maketitle

\begin{abstract}
We use machine learning for designing a medium frequency trading strategy for a portfolio of 5 year and 10 year US Treasury note futures. We formulate this as a classification problem where we predict the weekly direction of movement of the portfolio using features extracted from a deep belief network trained on technical indicators of the portfolio constituents. The experimentation shows that the resulting pipeline is effective in making a profitable trade.
\end{abstract}

\section{Introduction and Related work}
Machine learning application in finance is a challenging problem owing to low signal to noise ratio. Moreover, domain expertise is essential for engineering features which assist in solving an appropriate classification or regression problem. Most prior work in this area concentrates on using popular ML techniques like SVM \cite{huang2005forecasting}, \cite{tay2001application}, \cite{kim2003financial} and neural networks \cite{kaastra1996designing}, \cite{gately1995neural}, \cite{trippi1992neural}  coupled with rigorously designed features, and the general area of focus is financial time series forecasting. With deep learning techniques, we can learn the latent representation of the raw features and use this representation for further analysis \cite{erhan2010does}. In this paper, we construct a minimal risk portfolio of 5 year and 10 year T-note futures and use a machine learning pipeline to predict weekly direction of movement of the portfolio using features derived from a deep belief network. The prediction from the pipeline is then used in a day trading strategy. Using derivatives instead of the underlying entities themselves leads to a more feasible problem, since derivatives are less volatile and hence have clearer patterns.

The rest of the paper is divided into 4 sections. In section 2, we describe the dataset and the raw features. Section 3 discusses the methodology in rigour. In section 4, we highlight the findings of the experiment. Finally, we conclude with the possible improvements in this pipeline.

\section{Dataset and raw features}
We use data for 5 year and 10 year US treasury note futures (shortcodes ZF and ZN respectively) obtained from \emph{quandl}\footnote{https://www.quandl.com/data/CME}. We have daily data from 1985, where each data row consists of the opening price\footnote{https://www.cmegroup.com/education/files/understanding-treasury-futures.pdf}, the closing price, the high price, the low price, the open interest rate and the trading volume. Note that all these prices are mid-prices, hence the bid-ask spread is ignored. The prices on which we perform the rest of the analysis are obtained as the mid-price of the opening and the closing price each for ZF and ZN. This precludes us from trading at a higher frequency than placing a single trade in the day, but simplifies classification. Figure 1 shows the historical price evolution of ZF and ZN. As expected, in the long term the bond market goes up, but the relative spread has minimal drift.
\begin{figure}[!htb]
\centering
\includegraphics[scale=0.45]{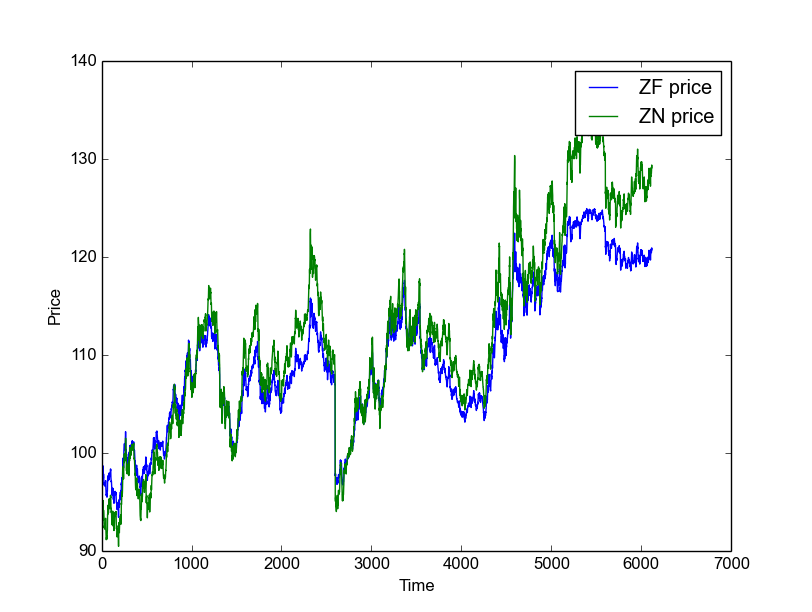}
\caption{Daily Instrument price series}
\label{fig:my_label}
\end{figure}

After removing rows containing missing information and some high leverage points during the 2008 financial crisis, we are left with around 6000 data points in total. We use 80\% of the data for training, the next 10\% data as validation set and the remaining 10\% data for testing purpose. We do not perform a $k$-fold validation as we are only interested in how a historical model performs on new data. The number of positive and negative examples (as defined below) are 2325 and 2608 for the training set, 280 and 310 for the validation set, and 295 and 315 for the test set.

Our raw features consist of daily trend (current price - moving average) computed over the past week with weekly and biweekly moving averages each for ZF and ZN. Hence, we have $5 \times 2 \times 2 = 20$ raw features. We perform a min-max normalization of these features for later analysis. With each data point, we associate a label +1 if the portfolio price (defined below) goes up 5 days from now and -1 otherwise.

\section{Methodology}
The flowchart below shows the complete ML pipeline of the strategy. A component wise description follows.
\vspace{4mm}
\begin{figure}[!htb]
\setlength{\unitlength}{0.14in} 
\begin{picture}(32,6) 
\put(5,4){\framebox(6,3){Filter}}
\put(16,4){\framebox(8,3){Portfolio}}
\put(5,-2){\framebox(6,3){Features}}
\put(4,-8){\framebox(7,3){Stacked RBMs}}
\put(17,-8){\framebox(6,3){Classifier}}
\put(2,6.5){\vector(1,0){3}}\put(11,6.5){\vector(1,0){5}}
\put(2,4.5){\vector(1,0){3}}\put(11,4.5){\vector(1,0){5}}
\put(2,7){$ZF_{raw}$}\put(2,5){$ZN_{raw}$}
\put(12,7){$ZF_{norm}$}\put(12,5){$ZN_{norm}$}
\put(7,4){\vector(0,-1){3}}\put(9,4){\vector(0,-1){3}}
\put(8,-2){\vector(0,-1){3}}
\put(15,0){$y_{pred}(k_1*ZF - k_2*ZN)$}
\put(8,-3.5){$f(ZF,ZN)$}
\put(19.5,4){\vector(0,-1){9}}
\put(11,-6.5){\vector(1,0){6}}
\put(11,-6){$f'(ZF,ZN)$}
\end{picture}
\vspace*{25mm}
\caption{ML pipeline of the strategy} 
\label{fig:lnlblock} 
\end{figure}
\subsection{Portfolio construction}
\begin{figure}[!htb]
\begin{center}
\includegraphics[scale=0.45]{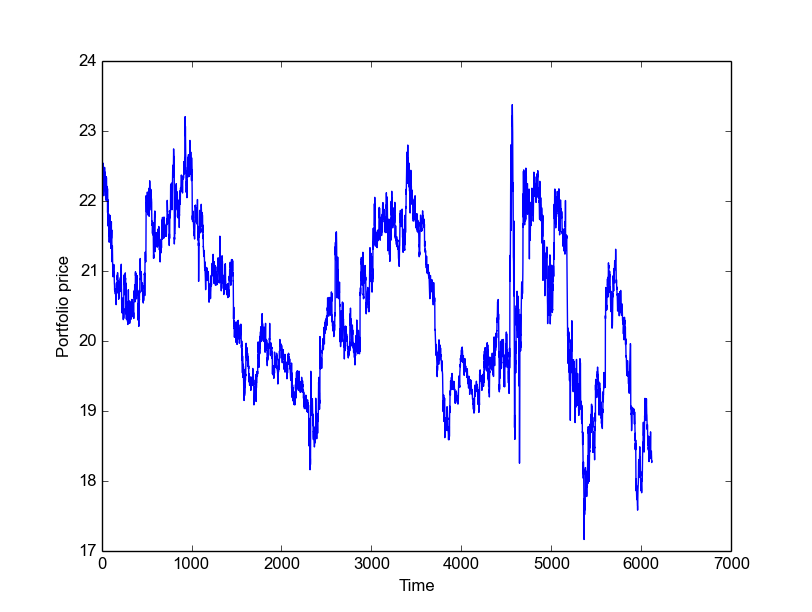}
\end{center}
\caption{Portfolio daily price series}
\label{fig:portprice}
\end{figure}

For obtaining the daily price series of the portfolio, we use principal component analysis (PCA) on the price series of ZF and ZN. PCA transforms data in a higher dimensional space to a space with equal or lower dimensionality with the basis vectors ordered by the percentage of variance explained by projection of the data along them. Since treasury note derivatives depend singularly on the interest rate, around 99\% of the variance in the data is explained by the first principal component. Hence, in practice, the loadings of the second principal component remain fairly constant when computed over a long duration of time. Moreover, one observes stationarity in the resultant portfolio price series, implying that the portfolio is minimal risk. Hence, $P_{port} = PC2_{5yr}*P_{5yr} + PC2_{10yr}*P_{10yr}$, where $PC2_{5yr}$ and $PC2_{10yr}$ are expected to hold opposite signs. Subsequently, whenever we perform a single unit transaction on the portfolio, we go long on one of the futures and short on the other in proportion to the magnitude of their loadings in the second component. This ensures that our positions in the two legs of the portfolio are hedged. The daily portfolio price series for the whole data is displayed in Figure \ref{fig:portprice}.  Observe that the drift component of the price series is almost negligible.
\subsection{Feature generation}
\begin{figure}[!htb]
\centering
\includegraphics[scale=0.4]{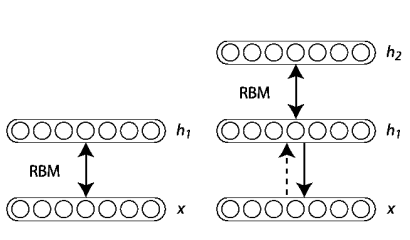}
\caption{Representative DBN}
\label{fig:representativeDBN}
\end{figure}

To extract latent representation of the raw features defined from the instruments, we use a deep belief network (DBN), which consists of stacked Restricted Boltzmann machines. A Boltzmann machine is an energy based model where the energy is a linear function of the free parameters\footnote{http://deeplearning.net/tutorial/rbm.html}. A Restricted Boltzmann machine (RBM) consists of a bipartite graph of a layer of visible units and a layer of hidden units, and can be used to learn a probability distribution over the input. Both layers of an RBM share weights, which are learned using contrastive divergence algorithm \cite{hinton2002training}. A DBN models $P(x, h^1, \ldots, h^{\ell}) = \left(\prod_{k=0}^{\ell-2} P(h^k|h^{k+1})\right) P(h^{\ell-1},h^{\ell})$, where $x=h^0, P(h^{k-1} | h^k)$ is a conditional distribution for the visible units conditioned on the hidden units of the RBM at level k, and $P(h^{\ell-1}, h^{\ell})$ is the visible-hidden joint distribution in the top-level RBM (see figure \ref{fig:representativeDBN})\footnote{Adopted from\\ http://www.iro.umontreal.ca/$\sim$ lisa/twiki/bin/view.cgi/Public/DeepBeliefNetworks }.  In our pipeline, the belief network consists of 2 RBMs where the first RBM is Gaussian-Bernoulli and the second RBM is Bernoulli. For best performance, both RBMs should be continuous (Gaussian is a special case), but training these RBMs is extremely difficult \cite{hinton2010practical}. In practice, using continuous input in only the first visible unit seems sufficient for our application. We use the standard algorithm proposed in \cite{hinton2006fast}, which consists of greedily doing contrastive divergence on each RBM for learning the reconstruction weight matrix and passing the transformed input to the next RBM. The output of DBN is a binary latent representation of the original input features derived from the corresponding Bernoulli distribution.
\subsection{Fine-tuning and classification}
The latent representation of the features is then used for constructing a classifier for the prediction goal defined above. There are three classifiers which we experiment with.
\begin{itemize}
\item Logistic regression - In logistic regression, we model $P(y_i|x_i) = \phi(x_i)^{\frac{1+y}{2}}(1-\phi(x_i))^{\frac{1-y}{2}}$ and $\phi(x_i) = \frac{1}{1+\exp(-\theta^T x_i)}$ and maximise the log likelihood of the input. To prevent over-fitting, typically a negative ridge penalty is also associated with the objective function. 
\item Support vector machine - An SVM discriminates between data by constructing a hyperplane $w^T\phi(x)+b = 0$ by minimising $\frac{||w||^2}{2}+C\sum \epsilon_i$ subject to $y_i(w^T \phi(x_i) + b) \geq 1 - \epsilon_i, \epsilon_i \geq 0\, \forall \, i$, where $\phi(x_i)$ is either $x_i$ or a higher dimensional representation of $x_i$. In either case, it does not need to be computed explicitly as $(w^T \phi(x) + b)$ can be obtained via ``kernel trick'' for classification since $w$ is a linear combination of a few $\phi(x_i)$ which are called support vectors.
\item Neural Network - A neural network consists of an input layer, a few hidden layer of nodes (typically 1 or 2) and an output layer with number of nodes equal to the cardinality of categories. It minimises $\displaystyle\sum_{i=1}^{m}(y_i - \hat{y_i})^2$, where $\hat{y_i} = h(g(x_i))$, $g$ is some linear combination of node values in the hidden layers of the network and $h$ is an activation function (typically sigmoid or hyperbolic tangent function). Training a neural network involves minimising the mean square error (MSE) defined above using gradient descent and back-propagating the gradient across multiple hidden layers for updating the weights between nodes. Here, instead of randomly initialising the neural network, we use the reconstruction weights learned in the deep belief network to form a neural network with 2 hidden layers.
\end{itemize}
\section{Experimentation and results}
\subsection{Portfolio weights}
We perform PCA on the training data by standardising the raw price series of ZN and ZN. We use the mean and standard deviation of the training to standardise the raw price series of validation and test for computing PCA on them separately. 
The second component loadings come out to be 0.83 for ZF and -0.59 for ZN in the training, 0.76 and -0.62 for the validation set, and 0.81 and -0.6 for the test. The validation set loading confirms the hypothesis that the portfolio is minimal risk. We use the training loading in training, validation and test sets for generating the labels for our supervised learning algorithms.
\subsection{Training the model}
There are two stages of training. In the first stage the DBN parameters are learned in an unsupervised manner and in the second stage, fine-tuning of the classifier parameters is performed using supervised learning.
\subsubsection{DBN training}
\begin{figure}[!htb]
\includegraphics[trim={4cm 1cm 3cm 1cm}, clip,width=0.5\textwidth,height=0.3\textheight]{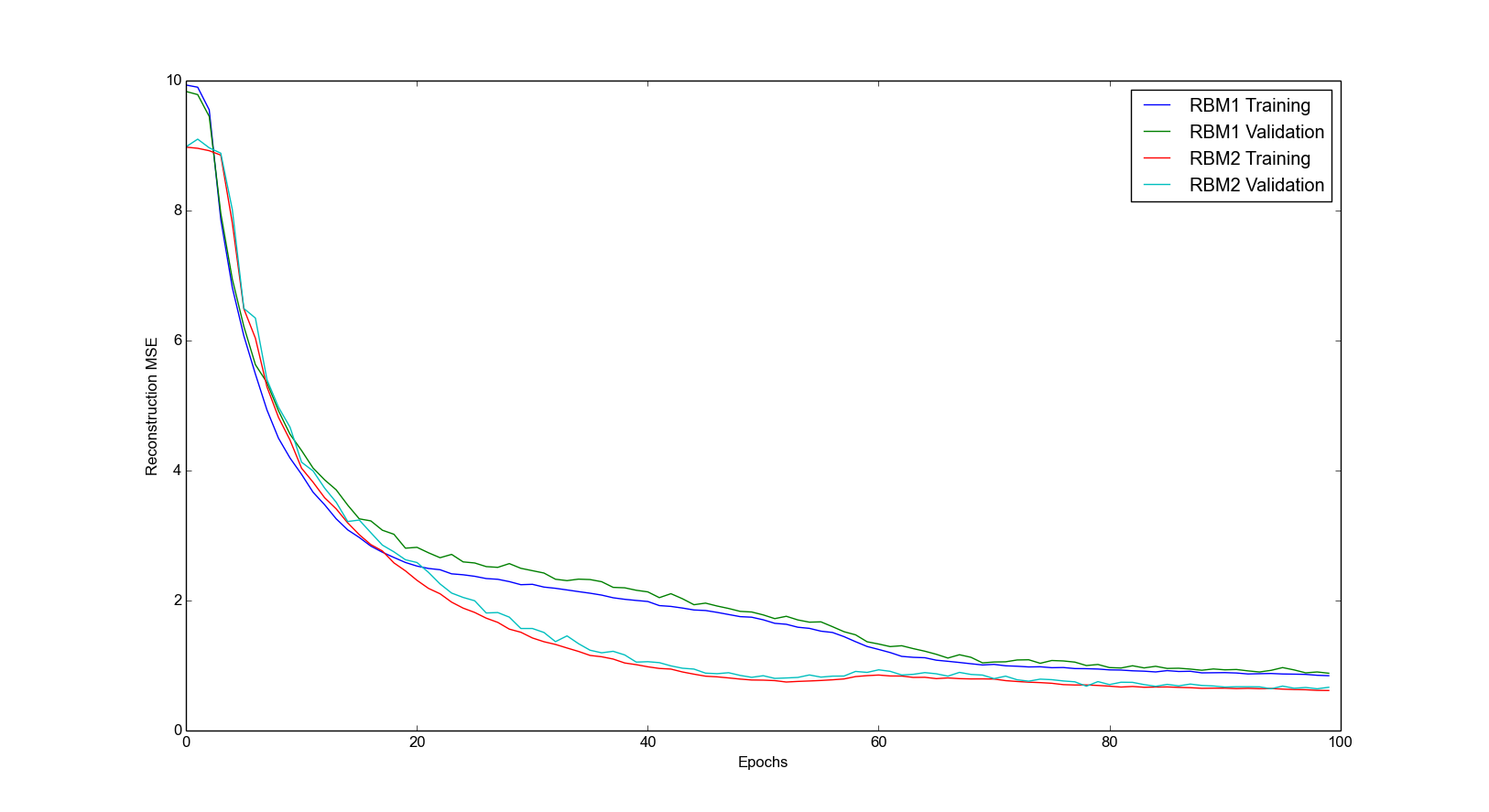}
\caption{Reconstruction MSE vs epochs for two stacked RBMs in the belief network}
\label{reconstructionerror}
\end{figure}

In our algorithm, the purpose of the belief network is to learn a hidden representation of the raw features used. Using recommendation from \cite{bengio2012practical}, we create a stack of 2 RBMs, with the first RBM containing 15 nodes in the hidden layer and the second RBM containing 20 nodes in the hidden layer. Each RBM is trained for 100 iterations with a mini-batch size of 100 in the contrastive divergence algorithm implemented in python. Since there is a tendency for the adjacent data points in the price series to be correlated, we shuffle the training data at the beginning of every macro-iteration. All hyper-parameters, including the node sizes in the hidden layers, are selected by monitoring the average reconstruction error on the validation set. The reconstruction error evolution for the best set of hyper-parameters on training and validation sets is shown in figure \ref{reconstructionerror}.
\subsubsection{Supervised learning}
\begin{figure}[!htb]
\centering
\includegraphics[width=0.5\textwidth]{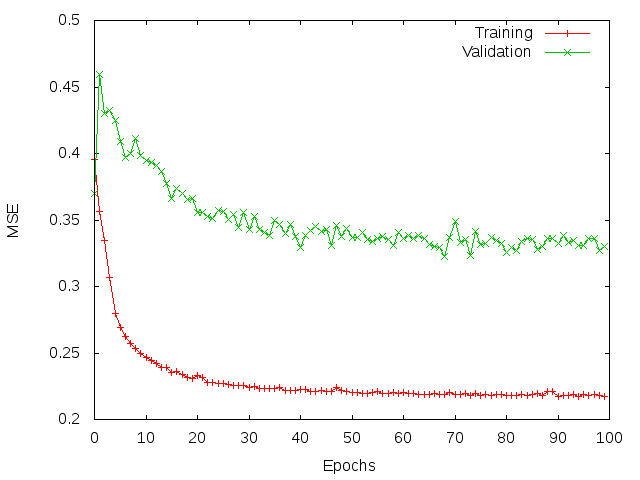}
\caption{MSE vs epochs for neural network}
\label{msevsepochs}
\end{figure}

We optimise the parameters for neural network by gradient descent and for logistic regression by gradient ascent. The optimal number of epochs used for the algorithm is decided by the performance on the validation set. For SVM, we use the validation set for optimising the cost parameter $C$. We use \emph{scikit} \cite{scikit-learn} implementation of SVM with a Gaussian Kernel, own implementation of logistic regression and implementation of \cite{IMM2012-06284} for neural network. An example evolution of MSE with epochs for neural network is shown in figure \ref{msevsepochs}.
\subsection{Predictions from the classifier}
\begin{figure}[!htb]
        \includegraphics[scale=0.45]{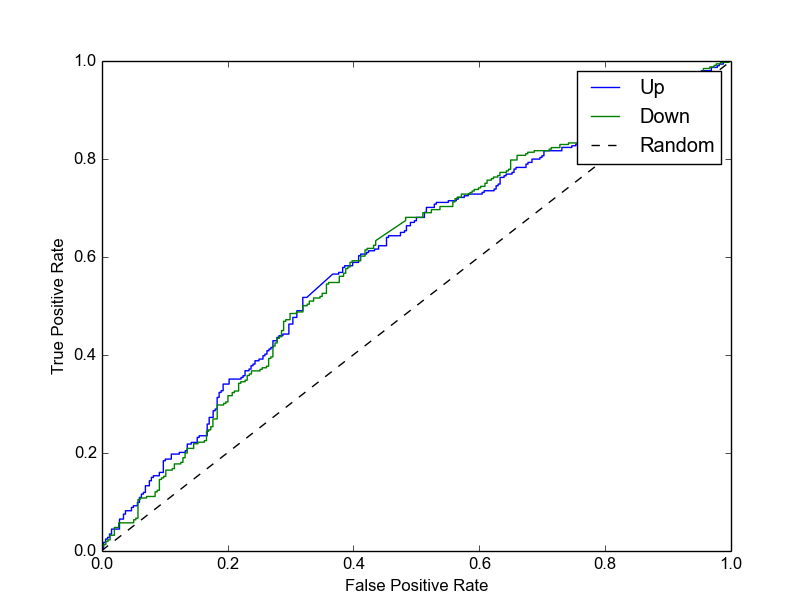}
        \caption{ROC curve for SVM}
        \label{fig:roc_svm}
\end{figure}

\begin{figure}[!htb]
        \includegraphics[scale=0.45]{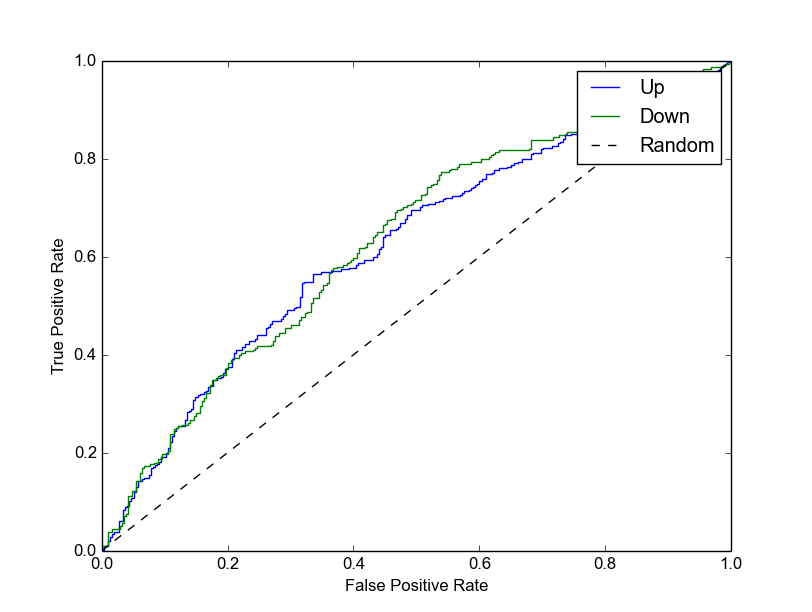}
        \caption{ROC curve for neural network}
        \label{fig:roc_nn}
\end{figure}

\begin{figure}[!htb]
        \centering
        \includegraphics[trim={3cm 0 3cm 2cm},clip,width=0.5\textwidth,height=0.3\textheight]{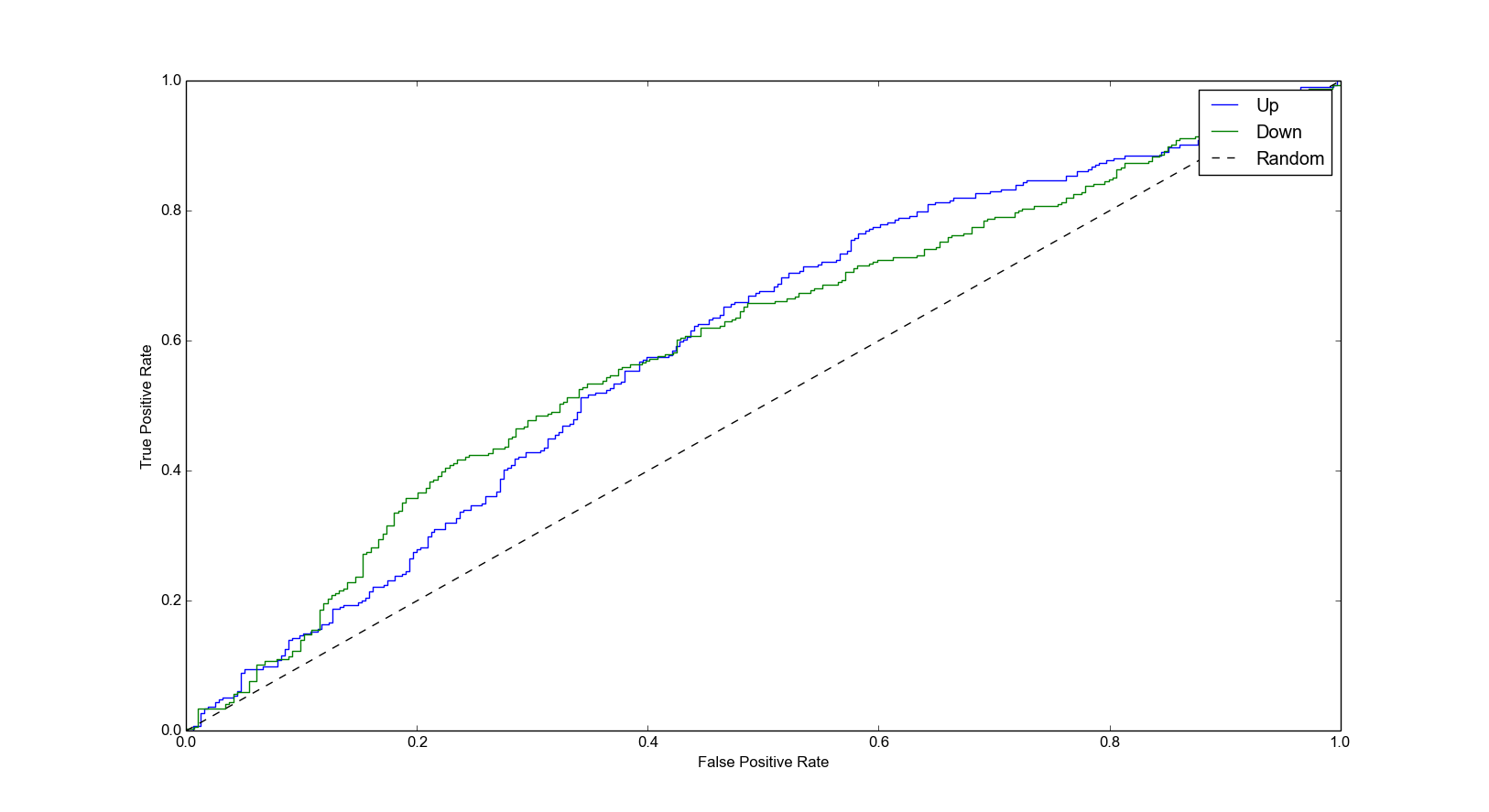}
        \caption{ROC curve for Logistic Regression}
        \label{fig:roc_lr}
\end{figure}

The precision and recall rates for the three classifiers in shown in Table 1 and Table 2 respectively for the test data and in Table 3 and Table 4 respectively for the training data. The corresponding ROC curves for test (True positive vs False positive rates) are shown in Figure \ref{fig:roc_svm}, \ref{fig:roc_nn} and \ref{fig:roc_lr}. For our application, we lay more emphasis on recall than precision since we do not want to lose any profitable trading opportunity that exists. The average recall hovers around 60\% for all three versions of classification in both training and test. This points at a non-trivial bias existing in the models. However, similar accuracies in training and test imply that the models can be generalisable enough. During training, we also observed high variance in the neural network parameters (refer figure \ref{msevsepochs} above) which we fix by introducing a momentum value of 0.5. This shrinks the existing values of the coefficients by half during every update of mini-batch gradient descent. Although the accuracy is not very high, in financial applications accuracies which are 5-10\% higher than random can guarantee a profitable trade.

\begin {table}[!htb]
\begin{center}
\begin{tabular}{ |l|c|c| } 
 \hline
 \textbf{Algorithm} & Actual $\Uparrow$ & Actual $\Downarrow$ \\ 
 \hline
 SVM & 60\% & 61.69\% \\ 
 Logistic Regression & 58.73\% & 62.37\% \\
 Neural Network & 57.77\% & 59.66\% \\ 
 \hline
\end{tabular}
\caption {Test Recall rate for each direction}
\end{center}
\end{table}
\begin{table}[!htb]
\begin{center}
\begin{tabular}{ |l|c|c| } 
 \hline
 \textbf{Algorithm} & Actual $\Uparrow$ & Actual $\Downarrow$ \\ 
 \hline
SVM & 62.58\% & 58.59\% \\ 
Logistic Regression & 62.5\% & 59.09\% \\
Neural Network & 60.46\% & 56.95\% \\ 
 \hline
\end{tabular}
\caption {Test Precision rate for each direction}
\end{center}
\end{table}
\begin {table}[!htb]
\begin{center}
\begin{tabular}{ |l|c|c| } 
 \hline
 \textbf{Algorithm} & Actual $\Uparrow$ & Actual $\Downarrow$ \\ 
 \hline
 SVM & 61.84\% & 58.72\% \\ 
 Logistic Regression & 60.51\% & 59.45\% \\
 Neural Network & 65.17\% & 54.26\% \\ 
 \hline
\end{tabular}
\caption {Training Recall rate for each direction}
\end{center}
\end{table}
\begin{table}[!htb]
\begin{center}
\begin{tabular}{ |l|c|c| } 
 \hline
 \textbf{Algorithm} & Actual $\Uparrow$ & Actual $\Downarrow$ \\ 
 \hline
SVM & 60.92\% & 59.65\% \\ 
Logistic Regression & 60.45\% & 58.86\% \\
Neural Network & 55.42\% & 66.35\% \\ 
 \hline
\end{tabular}
\caption {Training Precision rate for each direction}
\end{center}
\end{table}
\normalsize
\subsection{Trading strategy}
\begin{figure}[!htb]
\centering
\includegraphics[trim={3cm 1cm 3cm 2cm},clip,width=0.525\textwidth,height=0.3\textheight]{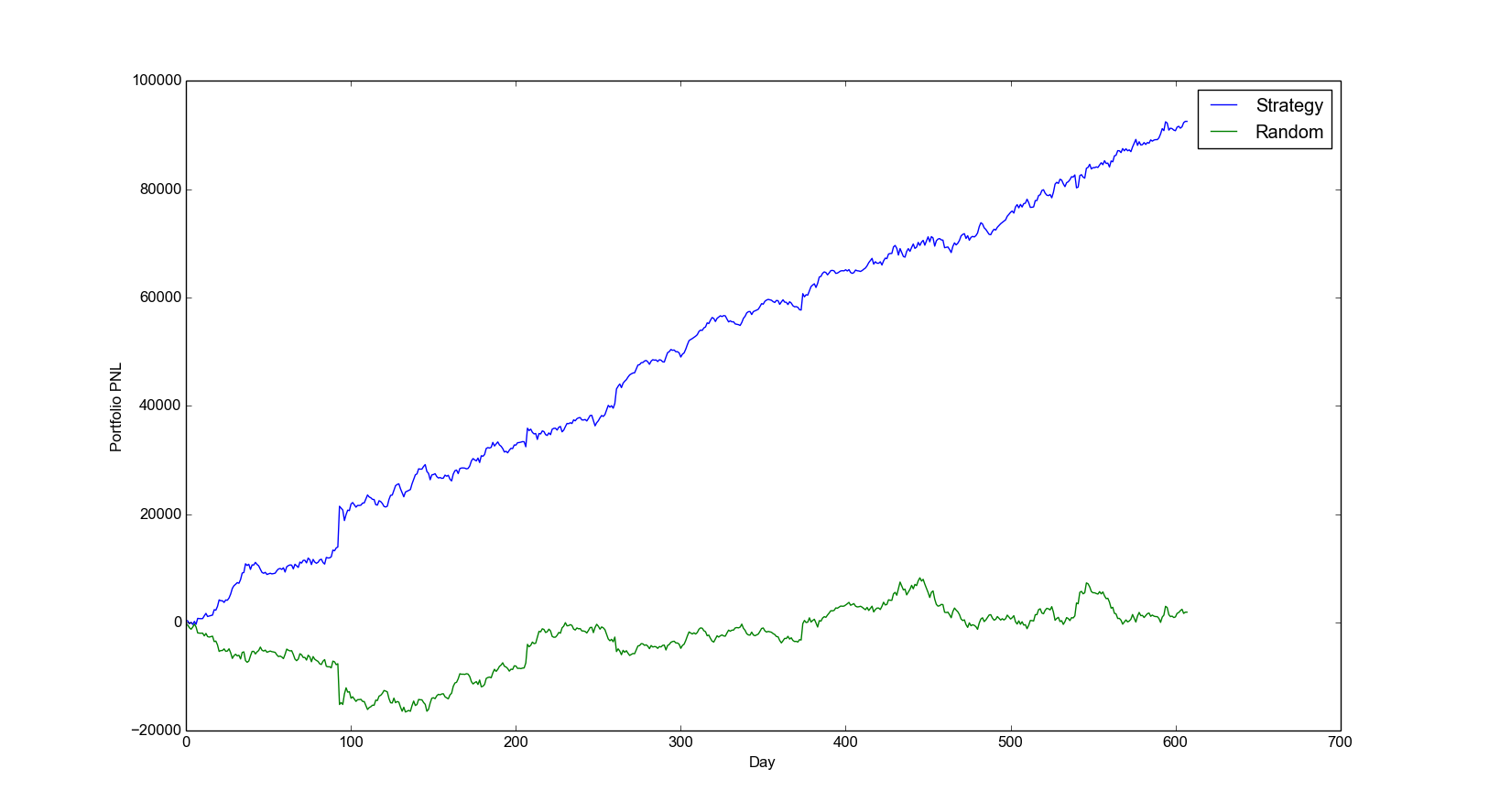}
\caption{Portfolio PNL vs number of days traded}
\label{fig:PNL}
\end{figure}

Using the prediction of the classifier, we build a day trading strategy where at the beginning of every day we go long (buy) on the portfolio if the price is going to go down after 5 days and short (sell) otherwise. We do not need to be long before we sell the portfolio since short selling is allowed on most commodity exchanges including CME. Since the portfolio is an artificial spread of the instruments, for going long we take a long position in ZF and a hedge adjusted short position in ZN and vice-versa for going short. The trade size that we use is 10 units of ZF and 8 units of ZN, which is the approximate integer ratio of the $PC2$ loadings defined earlier. Finally, at the end of 5 days, we square off the open position and accrue the resulting dollar tick difference in our PNL (Profit \& Loss). A qualitative comparison of this strategy against a strategy which places trades by obtaining predictions from a random classifier is shown in Figure \ref{fig:PNL}. Observe that the drawdown of the strategy is quite low, and excluding exchange transaction costs and taxes, the profitability for a trade size of 10 units for the portfolio is around 90000 dollars over a span of two and a half years.
\section{Conclusion and future work}
We build a day trading strategy for a portfolio of derivatives using features learned through a deep belief network from technical indicators of the constituents of the portfolio and using them to generate a predictor of weekly direction of movement of the portfolio. We observe that the feature reconstruction algorithm converges smoothly. We obtain a 5-10\% higher accuracy than a random predictor with the three classification techniques that we use and a much higher PNL. 

While a simplistic strategy can generate a moderately positive PNL, sophisticated algorithmic trading involves making more complex decisions for which one would want to quantify the magnitude of movement, for which building a regression based strategy can be helpful. For learning the quantified amount of portfolio price change, recurrent neural networks come in handy. Moreover, state of the art belief networks work very well for continuous inputs, and it is worthwhile to generate continuous features from RBMs for solving an appropriate classification or regression problem. The current approach for trading seems promising, and it would be interesting to use it for trading multiple spreads and butterflies of derivatives traded on multiple exchanges. Moreover, it would also be worthwhile to build a higher frequency trading strategy if one has access to tick by tick data.

\bibliography{mybib}{}

\begin{thebibliography}{10}

\bibitem{bengio2012practical}
Yoshua Bengio.
\newblock Practical recommendations for gradient-based training of deep
  architectures.
\newblock In {\em Neural Networks: Tricks of the Trade}, pages 437--478.
  Springer, 2012.

\bibitem{erhan2010does}
Dumitru Erhan, Yoshua Bengio, Aaron Courville, Pierre-Antoine Manzagol, Pascal
  Vincent, and Samy Bengio.
\newblock Why does unsupervised pre-training help deep learning?
\newblock {\em The Journal of Machine Learning Research}, 11:625--660, 2010.

\bibitem{gately1995neural}
Edward Gately.
\newblock {\em Neural networks for financial forecasting}.
\newblock John Wiley \& Sons, Inc., 1995.

\bibitem{hinton2010practical}
Geoffrey Hinton.
\newblock A practical guide to training restricted boltzmann machines.
\newblock {\em Momentum}, 9(1):926, 2010.

\bibitem{hinton2002training}
Geoffrey~E Hinton.
\newblock Training products of experts by minimizing contrastive divergence.
\newblock {\em Neural computation}, 14(8):1771--1800, 2002.

\bibitem{hinton2006fast}
Geoffrey~E Hinton, Simon Osindero, and Yee-Whye Teh.
\newblock A fast learning algorithm for deep belief nets.
\newblock {\em Neural computation}, 18(7):1527--1554, 2006.

\bibitem{huang2005forecasting}
Wei Huang, Yoshiteru Nakamori, and Shou-Yang Wang.
\newblock Forecasting stock market movement direction with support vector
  machine.
\newblock {\em Computers \& Operations Research}, 32(10):2513--2522, 2005.

\bibitem{kaastra1996designing}
Iebeling Kaastra and Milton Boyd.
\newblock Designing a neural network for forecasting financial and economic
  time series.
\newblock {\em Neurocomputing}, 10(3):215--236, 1996.

\bibitem{kim2003financial}
Kyoung-jae Kim.
\newblock Financial time series forecasting using support vector machines.
\newblock {\em Neurocomputing}, 55(1):307--319, 2003.

\bibitem{IMM2012-06284}
R.~B. Palm.
\newblock Prediction as a candidate for learning deep hierarchical models of
  data.
\newblock Master's thesis, 2012.

\bibitem{scikit-learn}
F.~Pedregosa, G.~Varoquaux, A.~Gramfort, V.~Michel, B.~Thirion, O.~Grisel,
  M.~Blondel, P.~Prettenhofer, R.~Weiss, V.~Dubourg, J.~Vanderplas, A.~Passos,
  D.~Cournapeau, M.~Brucher, M.~Perrot, and E.~Duchesnay.
\newblock Scikit-learn: Machine learning in {P}ython.
\newblock {\em Journal of Machine Learning Research}, 12:2825--2830, 2011.

\bibitem{tay2001application}
Francis~EH Tay and Lijuan Cao.
\newblock Application of support vector machines in financial time series
  forecasting.
\newblock {\em Omega}, 29(4):309--317, 2001.

\bibitem{trippi1992neural}
Robert~R Trippi and Efraim Turban.
\newblock {\em Neural networks in finance and investing: Using artificial
  intelligence to improve real world performance}.
\newblock McGraw-Hill, Inc., 1992.

\end{thebibliography}
\bibliographystyle{plain}

\end{document}